\def\del{\partial}
\def\w{\omega}
\def\R{\tilde{R}}
\def\e{\xi}
\def\r1{r_{+}}
\def\r2{r_{-}}

\documentclass[11pt,a4paper,english,amssymb,nofootinbib,superscriptaddress,showpacs]{revtex4}
\usepackage{lmodern}

\usepackage[T1]{fontenc}
\usepackage[latin9]{inputenc}
\usepackage{babel}

\usepackage{amsmath}
\usepackage{graphicx}
\usepackage{amssymb}
\usepackage{esint}
\usepackage[unicode=true, pdfusetitle,
 bookmarks=true,bookmarksnumbered=false,bookmarksopen=false,
 breaklinks=false,pdfborder={0 0 1},backref=false,colorlinks=false]
 {hyperref}

\makeatletter
\@ifundefined{textcolor}{}
{%
 \definecolor{BLACK}{gray}{0}
 \definecolor{WHITE}{gray}{1}
 \definecolor{RED}{rgb}{1,0,0}
 \definecolor{GREEN}{rgb}{0,1,0}
 \definecolor{BLUE}{rgb}{0,0,1}
 \definecolor{CYAN}{cmyk}{1,0,0,0}
 \definecolor{MAGENTA}{cmyk}{0,1,0,0}
 \definecolor{YELLOW}{cmyk}{0,0,1,0}
 }

\usepackage{latexsym}\usepackage{bm}

\makeatother

\makeatother

\begin{document}

\title{ Covariant Symplectic Structure and Conserved Charges of Topologically Massive Gravity   }

\author{ Caner Nazaroglu}

\email{e155529@metu.edu.tr}

\affiliation{Department of Physics,\\
 Middle East Technical University, 06531, Ankara, Turkey}

\author{Yavuz Nutku}

\affiliation{Feza G\"ursey Institute P.O.Box 6 \c{C}engelk\"oy, Istanbul 81220 Turkey}

\author{Bayram Tekin}

\email{btekin@metu.edu.tr}

\affiliation{Department of Physics,\\
 Middle East Technical University, 06531, Ankara, Turkey}

\date{\today}
\begin{abstract}
 We present the covariant symplectic structure of the
Topologically Massive Gravity and find a compact expression for the conserved charges of generic spacetimes with 
 Killing symmetries.
 \tableofcontents{}\[
\]

\end{abstract}
\maketitle

\section{Introduction}

The classical theory of symplectic structure was cast into
covariant form by Gotay\footnote{One of us (YN) has written a
number of papers on covariant symplectic structure without being
aware of Gotay's paper, principally because it was published in a
conference proceedings. Thanks are due to Dr. Partha Gupta for
calling attention to Gotay's work.} \cite{g}, Witten \cite{w} and
Zuckerman \cite{z} almost simultaneously. This is an important
development in the theory of symplectic structure founded
in the latter part of the nineteenth century . It is particularly
apt for dealing with covariant field theories such as colored
gauge theories and gravity.

The covariant symplectic structure is defined by a
vector-density-valued $2$-form which is closed and divergence
free. Its time component agrees with the inverted Dirac \cite{d}
bracket, {\it i.e.} the Poisson bracket of the constraints.

For Yang-Mills and Einstein theories the covariant symplectic
structure was given in \cite{w}. These results can directly be
carried over into Kaluza-Klein theories in higher dimensions.
However, in $4n+3$ dimensions there is an added feature, namely
the existence of Chern-Simons invariants. In particular for the
simplest case of three dimensions we have the theories of
Yang-Mills-Chern-Simons, as well as the Topologically Massive Gravity
(TMG) \cite{djt}. TMG is a higher derivative dynamical theory with a 
single degree of freedom and compared to the pure Einstein's gravity in three dimensions, its local 
structure is quite rich. For this theory the Dirac
constraint analysis was carried out by Deser and Xiang \cite{dx}. 
In this paper, to define the classical phase space in a covariant way, 
we find the symplectic structure. The analysis naturally gives the conserved charges that are generated by
Killing symmetries for generic spacetimes. 
 
The lay out of the paper is as follows: In Section II, we give the symplectic structure of the Maxwell-Chern-Simons theory as a warm-up problem. In Section III,  we find the symplectic 2-form for TMG and show that it is conserved and closed. In section IV, we study the diffeomorphism invariance of the symplectic structure and show that it has vanishing components in the gauge directions. In that section, we also find the conserved charges and compute the energy of the BTZ black hole.

\section{Maxwell-Chern-Simons theory}

Before we find the symplectic 2-form of TMG,  we start with a simpler model, that is the  topologically massive electrodynamics. The existence of the symplectic structure leads to a covariant canonical description of the classical and quantum theory.   Since one must choose 
a time and  define momenta { \it etc .}  in a canonical description, it might appear that one cannot have covariance. But, as was shown in  \cite{g,w,z},  all the properties of the phase space ($Z$) is encoded in the symplectic structure and momenta need not be defined.  We will use the notation and follow the construction of \cite{w}. On $Z$ one defines a 2-form $\omega$ which is closed ( $\delta \omega=0$ ) and non-degenerate (save for the gauge directions); {\it i.e.}, for any vector field $v$ on $Z$, if $\iota _v  \omega = v^I \omega_{ I J}=0$, then $v=0$ (which just means, as a matrix $\omega$ has no zero eigenvalues and hence it is invertible).  What is quite remarkable is that, in local 
coordinates of the phase space ($q^I$), the basic Poisson bracket is given by the components of the inverse of the 2-form $\{ q^I, q^J\} =\omega^{I J}$.  Therefore, one can use the symplectic 2-form to carry out a covariant, geometric quantization  of the system. 

The Lagrangian of the Maxwell-Chern-Simons theory is given as
\begin{equation}
 {\cal L} = -\frac{1}{4} F_{\mu \nu} F^{\mu \nu} 
 + \kappa \epsilon^{\mu \nu \lambda} F_{\mu \nu} A_\lambda  \label{lem},
\end{equation}
where $\kappa$ is a coupling constant. From the first variation of
the action we obtain the field equation
\begin{equation}
\del_\mu F^{\mu \nu} + 2\kappa \epsilon^{\mu \lambda \nu} F_{\mu \lambda} = 0,
\label{csem}
\end{equation}
along with the boundary term
\begin{equation}
\alpha^\mu = - F^{\mu \nu} \delta A_\nu - 2 \kappa \epsilon^{\mu \nu \lambda} A_\nu \delta A_\lambda.
\end{equation} 
Then we obtain the symplectic current as
\begin{eqnarray}
J^\mu = - \delta \alpha^\mu 
=  \delta F^{\mu \nu} \wedge \delta A_\nu + 2 \kappa  \epsilon^{\mu \nu \lambda} 
\delta A_\nu \wedge \delta A_\lambda \label{emom}.
\end{eqnarray}
It is easy to see from there, that $J^\mu$ is conserved on shell and closed
\begin{equation}
\del_\mu J^\mu = 0 \textrm{, \ \ \  } \delta J^\mu = 0. \label{closure}
\end{equation}
Therefore the two-form defined as 
\begin{equation} 
\w=\int_\Sigma d \Sigma_\mu J^\mu = \int_\Sigma d \Sigma_\mu \left( \delta F^{\mu \nu} \wedge \delta A_\nu + 2 \kappa  \epsilon^{\mu \nu \lambda}
\delta A_\nu \wedge \delta A_\lambda \right),
\end{equation}
where $\Sigma$ is 
an initial value hypersurface, is closed and Poincar\'e invariant and hence gives the symplectic
structure we seek on the space of classical solutions (let us call it $\hat{Z}$), however we still have to show that it is 
a gauge invariant closed two-form in the {\it quotient} of the solution space, $Z = \hat{Z}/G$, where $G$ is the 
group of gauge transformations, which, in this case, is $U(1)$. Showing the gauge invariance of $\w$ on the full space
of solutions is easy, since under infinitesimal gauge transformations we have
\begin{equation}
A_\lambda \rightarrow A_\lambda + \del_\lambda \e  \textrm{ \ \ \  implying \ \ \    } \delta A_\lambda \rightarrow \delta A_\lambda \mbox{ and } 
 \delta F_{\mu \nu} \rightarrow \delta F_{\mu \nu}.
\end{equation}
Hence, $\w$ is gauge invariant on the full solution space. Let us now show that it is also gauge invariant on $Z$. To do this we must show that $\w$
has vanishing components along the pure gauge directions. Then, we can split the field into pure gauge and non-gauge parts as
\begin{equation}
\delta A'_\mu = \delta A_\mu + \del_\mu \e,
\end{equation}
where $\e$ is a one form on the cotangent space of the phase space manifold. Then the change in $\w$ due to this pure gauge part is
\begin{equation}
\Delta \w = \int_\Sigma d \Sigma_\mu \left( \del_\lambda \delta F^{\lambda \mu} + 4 \kappa \epsilon^{\lambda \nu \mu} \del_\lambda \delta A_\nu \right) \wedge \e
		+   \int_\Sigma d \Sigma_\mu \del_\lambda \left[ \left( \delta F^{\mu \lambda} + 4 \kappa \epsilon^{\mu \nu \lambda} \delta A_\nu  \right) \wedge \e \right].
\end{equation}
The first term vanishes on shell and the second one is a boundary term disappearing for fields decaying sufficiently fast. Therefore, $\w$ is the sought-after symplectic
structure on the classical phase space modulo gauge transformations.

\section{Topologically massive gravity}

For gravity the situation is slightly more complicated. The
action for TMG is given by \cite{djt}
\begin{equation}
 I =\int d^3 x \left[ \sqrt{g} R + \frac{1}{2 \mu}  \epsilon^{\alpha \beta \gamma}
 \Gamma^\mu_{\ \alpha \nu}  \left( \del_\beta \Gamma^\nu_{\ \gamma \mu} + \frac{2}{3}
  \Gamma^\nu_{\ \beta \rho} \Gamma^\rho_{\ \gamma \mu}  \right) \right],
 \label{lg}
\end{equation}
where $\epsilon^{\alpha \beta \gamma}$ is the totally antisymmetric Levi-Civita symbol, which,
as a tensor density, has the same weight as $\sqrt{g}$. (One can add a cosmological constant to this action; but this will not change the discussion below.)

Now, we calculate the variation of this action with respect to the metric:
\begin{equation}
\delta I = \delta I_{EH} + \delta I_{CS}.
\end{equation}
The Einstein-Hilbert term is known to yield
\begin{eqnarray}
\delta I_{EH} &=& \delta \int d^3 x \sqrt{g} R \nonumber \\ 
					&=& \int d^3 x \sqrt{g} \delta g^{\mu \nu} G_{\mu \nu} +
						 \int d^3 x \del_\alpha \left( \sqrt{g} g^{\mu \nu} \delta \Gamma^\alpha_{\ \mu\nu}
						 	- \sqrt{g} g^{\alpha \mu} \delta \Gamma^\nu_{\ \mu\nu} \right).
\end{eqnarray}
For Chern-Simons term we have
\begin{eqnarray}
\delta I_{CS} &=& \delta \int d^3 x \frac{1}{2 \mu}  \epsilon^{\alpha \beta \gamma}
 \Gamma^\mu_{\ \alpha \nu}  \left( \del_\beta \Gamma^\nu_{\ \gamma \mu} + \frac{2}{3}
  \Gamma^\nu_{\ \beta \rho} \Gamma^\rho_{\ \gamma \mu}  \right) \nonumber \\
      &=& \frac{1}{2 \mu} \int d^3 x \epsilon^{\alpha \beta \gamma} \delta 
  			\Gamma^\mu_{\ \alpha\nu} R^\nu_{\ \mu\beta\gamma} + \int d^3 x \del_\alpha 
  			\left( -\frac{1}{2\mu} \epsilon^{\alpha \nu \sigma} \Gamma^\rho_{\ \nu \beta}
  				\delta \Gamma^\beta_{\ \sigma \rho}  \right) \nonumber \\
  				&=& \frac{1}{\mu} \int d^3 x \sqrt{g} \delta g^{\mu \nu} C_{\mu \nu}
  				 + \int d^3 x \del_\alpha 
  			\left[ -\frac{1}{\mu} \epsilon^{\alpha \nu \sigma} \left(\tilde{R}^\rho_{\ \sigma} 
  			\delta g_{\nu \rho} + \frac{1}{2} \Gamma^\rho_{\ \nu \beta}
  				\delta \Gamma^\beta_{\ \sigma \rho}  \right) \right],
\end{eqnarray}
where the Cotton tensor is $C^{\mu \nu} = \frac{\epsilon^{\mu \beta \gamma}}{\sqrt{g}} \nabla_\beta \tilde{R}^\nu_{\ \gamma}$ with 
$\tilde{R}_{\mu \nu} = R_{\mu\nu} - \frac{1}{4} g_{\mu\nu} R$.
This yields the equation of motion $G_{\mu \nu} + \frac{1}{\mu} C_{\mu \nu} = 0$ and the boundary term
\begin{eqnarray}
\Lambda^\alpha = \Lambda^\alpha_{EH} + \Lambda^\alpha_{CS},
\end{eqnarray}
where
\begin{eqnarray}
 \Lambda^\alpha_{EH} &=&  \sqrt{g} g^{\mu \nu} \delta \Gamma^\alpha_{\ \mu\nu}
						 	- \sqrt{g} g^{\alpha \mu} \delta \Gamma^\nu_{\ \mu\nu}, \\
 \Lambda^\alpha_{CS}& =&   -\frac{1}{\mu} \epsilon^{\alpha \nu \sigma} \left( \tilde{R}^\rho_{\ \sigma} 
  			\delta g_{\nu \rho} + \frac{1}{2} \Gamma^\rho_{\ \nu \beta}
  				\delta \Gamma^\beta_{\ \sigma \rho}  \right).	
\end{eqnarray}

From the boundary terms one can construct the symplectic current as follows:
\begin{eqnarray}
J^\alpha = J^\alpha_{EH} + J^\alpha_{CS},
\end{eqnarray}
where
\begin{eqnarray}
 J^\alpha_{EH}  = - \frac{\delta \Lambda^\alpha_{EH}}{\sqrt{g}}
	= \delta  \Gamma^\alpha_{\ \mu\nu} \wedge  \left( \delta  g^{\mu \nu} + \frac{1}{2}  g^{\mu \nu} \delta \ln g \right)
	-  \delta  \Gamma^\nu_{\ \mu\nu} \wedge  \left( \delta  g^{\alpha \mu} + \frac{1}{2}  g^{\alpha \mu} \delta \ln g \right)
\end{eqnarray}
and
\begin{eqnarray}
J^\alpha_{CS} = - \frac{\delta \Lambda^\alpha_{CS}}{\sqrt{g}} 
	= \frac{1}{\mu} \frac{ \epsilon^{\alpha \nu \sigma}}{\sqrt{g}} \left( \delta  \tilde{R}^\rho_{\ \sigma} \wedge \delta g_{\nu \rho}
					+ \frac{1}{2} \delta \Gamma^\rho_{\ \nu \beta} \wedge  \delta \Gamma^\beta_{\ \sigma \rho}  \right).
\end{eqnarray}

Then, the symplectic two-form on the phase space of TMG, $\w=\int_\Sigma d \Sigma_\alpha \sqrt{g} J^{\alpha}$  reads 
\begin{equation}
\begin{aligned}
\w = \int_\Sigma d \Sigma_\alpha \sqrt{g} \Bigg[  &\delta  \Gamma^\alpha_{\ \mu\nu} \wedge  \left( \delta  g^{\mu \nu} + \frac{1}{2}  g^{\mu \nu} \delta \ln g \right)
	-  \delta  \Gamma^\nu_{\ \mu\nu} \wedge  \left( \delta  g^{\alpha \mu} + \frac{1}{2}  g^{\alpha \mu} \delta \ln g \right) \\
		&+   \frac{1}{\mu} \frac{ \epsilon^{\alpha \nu \sigma}}{\sqrt{g}} \left( \delta  \tilde{R}^\rho_{\ \sigma} \wedge \delta g_{\nu \rho}
					+ \frac{1}{2} \delta \Gamma^\rho_{\ \nu \beta} \wedge  \delta \Gamma^\beta_{\ \sigma \rho}  \right) \Bigg]     .
\label{symplectic_two}
\end{aligned}
\end{equation}
Without the use of field equations it is not difficult to see that the two-form is closed, $\delta \w = 0$.

The next part of the computation is to show that $J^\alpha$ is conserved on shell, that is, $\nabla_\alpha J^\alpha = 0$ modulo field equations 
and their variations, $\delta G_{\mu \nu} + \frac{1}{\mu} \delta C_{\mu \nu}= 0$. Below we give some details of this computation. Let us define
the covariant divergence of the current as
\begin{equation}
\nabla_\alpha J^\alpha \equiv I_1 + I_2 + \frac{1}{\mu} I_3,
\end{equation}
where
\begin{equation}
I_1 \equiv  \frac{1}{2} \nabla_\alpha \left(    g^{\mu \nu}  \delta  \Gamma^\alpha_{\ \mu\nu} \wedge \delta \ln g 
	-    g^{\alpha \mu} \delta  \Gamma^\nu_{\ \mu\nu} \wedge   \delta \ln g \right),
\end{equation}
\begin{equation}
I_2 \equiv  \nabla_\alpha \left(  \delta  \Gamma^\alpha_{\ \mu\nu} \wedge \delta  g^{\mu \nu}
	-  \delta  \Gamma^\nu_{\ \mu\nu} \wedge   \delta  g^{\alpha \mu} \right),
\end{equation}
and
\begin{equation}
I_3 \equiv  \nabla_\alpha \left[  \frac{ \epsilon^{\alpha \nu \sigma}}{\sqrt{g}} \left( \delta  \tilde{R}^\rho_{\ \sigma} \wedge \delta g_{\nu \rho}
					+ \frac{1}{2} \delta \Gamma^\rho_{\ \nu \beta} \wedge  \delta \Gamma^\beta_{\ \sigma \rho}  \right) \right].
\end{equation}

Using the Palatini identity, $\delta R_{\mu \nu} = \nabla_\alpha \delta \Gamma^\alpha_{\ \mu\nu} - \nabla_\mu \delta \Gamma^\alpha_{\ \nu\alpha}$ and the explicit form of 
$\delta\Gamma$ in terms of the metric and the symmetries of the involved tensors one can reduce $I_1$ and $I_2$ to the following forms:
\begin{eqnarray}
I_1 &=& \frac{1}{2} g^{\mu \nu} \delta R_{\mu \nu} \wedge \delta \ln g +  g^{\mu \nu}  \delta  \Gamma^\alpha_{\ \mu\nu} \wedge  \delta  \Gamma^\lambda_{\ \alpha\lambda},\\
I_2&=&  \delta R_{\mu \nu} \wedge \delta g^{\mu \nu}  -  g^{\mu \nu}  \delta  \Gamma^\alpha_{\ \mu\nu} \wedge  \delta  \Gamma^\lambda_{\ \alpha\lambda}.
\end{eqnarray}

With the help of field equations $I_1+I_2$ can be reduced to
\begin{equation}
I_1 + I_2 = \frac{1}{\mu} I_4
\end{equation}
where
\begin{equation}
I_4 = \delta C^{\mu \nu} \wedge  \left(   \delta g_{\mu \nu } -  \frac{1}{2} g_{\mu \nu} \delta \ln g  \right) -  C^{\mu \nu} \delta  g_{\mu \nu } \wedge  \delta \ln g.
\end{equation}

The variation of the Cotton tensor,
\begin{equation}
\delta  C^{\mu \nu} =  \frac{\epsilon^{\mu \beta \gamma}}{\sqrt{g}} \left( -\frac{1}{2}  \nabla_\beta \tilde{R}^\nu_{\ \gamma} \delta  \ln g + \nabla_\beta \delta \R^\nu_{\ \gamma}
	+ \R^\sigma_{\ \gamma} \delta \Gamma^\nu_{\ \beta\sigma} \right),
\end{equation}
can be used to reduce $I_4$ to 
\begin{equation}
I_4 =  \frac{\epsilon^{\mu \beta \gamma}}{\sqrt{g}} \left(  \R^\sigma_{\ \gamma} \delta  \Gamma^\nu_{\ \beta\sigma}
		+ \nabla_\beta \delta  \R^\nu_{\ \gamma}  \right) \wedge \delta  g_{\mu \nu} .
\end{equation}

$I_3$ can be brought into the form
\begin{equation}
I_3 = -  \frac{\epsilon^{\mu \beta \gamma}}{\sqrt{g}} \left( \nabla_\beta \delta \R^\nu_{\ \gamma} \wedge \delta  g_{\mu \nu}  + g_{\lambda \mu}  \delta \R^\nu_{\ \gamma}
		\wedge \delta  \Gamma^\lambda_{\ \beta\nu} +  \delta  \Gamma^\nu_{\ \mu\sigma}  \wedge \nabla_\beta  \delta  \Gamma^\sigma_{\ \gamma\nu} \right).
\end{equation}

Then combining $I_3$ and $I_4$ one obtains
\begin{equation}
\nabla_\alpha J^\alpha =  \frac{\epsilon^{\mu \beta \gamma}}{\mu \sqrt{g}} \delta \Gamma^\nu_{\ \beta\sigma} \wedge \left[    \delta   \left( g_{\mu \nu}  \R^\sigma_{\ \gamma}
		\right) + \nabla_\mu \delta \Gamma^\sigma_{\ \gamma\nu}  \right].
\label{conserved}
\end{equation}

Finally, using the explicit form of the Riemann tensor in terms of the connection and the three dimensional identities
\begin{eqnarray}
\epsilon^{\mu \beta \gamma} R^\sigma_{\ \mu\gamma\nu} &=& \epsilon^{\mu \beta \gamma} \left( \delta^\sigma_{\ \gamma} \R_{\mu\nu} + \R^\sigma_{\ \gamma} g_{\mu\nu} \right), \\
\epsilon^{\mu \beta \gamma}\delta R^\sigma_{\ \mu\gamma\nu} &=&  \epsilon^{\mu \beta \gamma} \delta^\sigma_{\ \gamma} \delta\R_{\mu\nu} 
											+ \epsilon^{\mu \beta \gamma} \delta \left( \R^\sigma_{\ \gamma} g_{\mu\nu} \right),
\end{eqnarray}
one can show that the right hand side of (\ref{conserved}) is zero and the symplectic current, $J^\alpha$, is covariantly conserved on shell.

Finally, we have to show that $\w$ is diffeomorphism invariant both in the full solution space and in the more relevant quotient space of solutions modulo the diffeomorphism group. 
The former computation requires no work since the constructed symplectic current only involves tensors as ingredients.
The latter one, on the other hand, is somewhat nontrivial, but it is quite fruitful since it will also give us the conserved charges corresponding to the Killing symmetries. Therefore, we devote the following section to this computation.

\section{Diffeomorphisms and conserved quantities}

To see that $\w$ has vanishing components in the pure gauge directions let us decompose the variation of the metric into non-gauge and pure gauge  parts:
\begin{equation}
\delta g'_{\mu \nu} = \delta g_{\mu \nu} + \nabla_\mu \e_\nu + \nabla_\nu \e_\mu,
\label{diffeo}
\end{equation}
where $\e$ is a one-form on the cotangent space of the phase space. Under this decomposition the relevant tensors split 
as:~\footnote{Note that the computation boils down to finding the Lie derivative of the associated tensors, $\mathcal{L}_\e T$, with respect to the vector $\e$.}
\begin{equation}
\delta \Gamma'^\lambda_{\ \mu \nu} = \delta \Gamma^\lambda_{\ \mu \nu} + \nabla_\mu \nabla_\nu \e ^\lambda + R^{\ \lambda}_{\nu \ \mu \beta} \e^\beta,
\end{equation}
\begin{equation}
\delta \tilde{R}'^\mu_{\ \nu} = \delta\tilde{R}^\mu_{\ \nu} +  \e^\beta \nabla_\beta  \R^\mu_{\ \nu}    + \R^\mu_{\ \beta} \nabla_\nu \e^\beta - \R_{\nu \beta} \nabla^\beta \e^\mu .
\end{equation}

The change in the symplectic current of the Einstein-Hilbert part can be computed as \cite{w}:
\begin{equation}
\Delta J_{EH}^\alpha = \nabla_\mu X_{EH}^{\mu \alpha} + R^{\mu \alpha} \left( \e_\mu \wedge \delta \ln g + 2 \e^\nu \wedge \delta g_{\mu \nu} \right)
		+ R^{\mu \nu} \delta g_{\mu \nu} \wedge \e^\alpha + \delta R \wedge  \e^\alpha + 2 \e_\mu \wedge \delta R^{\alpha \mu},
\label{gauge_einstein}
\end{equation}
where the $ X_{EH}^{\mu \alpha}$ is an antisymmetric tensor defined as:
\begin{equation}
 X_{EH}^{\mu \alpha} = \nabla^\mu \delta g^{\nu \alpha} \wedge \e_\nu +  \delta g^{\nu \alpha}\wedge \nabla_\nu  \e^\mu  + \frac{1}{2} \delta \ln g \wedge  \nabla^\alpha  \e^\mu
		+  \nabla_\nu \delta g^{\mu \nu} \wedge \e^\alpha + \nabla^\mu \delta\ln g \wedge \e^\alpha - \left( \alpha \leftrightarrow \mu \right).
\end{equation}

If we consider the pure Einstein-Hilbert theory alone, then the last four terms of (\ref{gauge_einstein}) vanish on shell and the first term is a boundary term, which vanishes for 
sufficiently decaying metric variations. Therefore, the corresponding symplectic two-form $\w_{EH}$ is diffeomorphism invariant on $Z$, the quotient space of classical solutions to the
diffeomorphism group. 

Let us now consider the change in the Chern-Simons part of the symplectic current:
\begin{equation}
\begin{aligned}
\mu \Delta J_{CS}^\alpha = \frac{\epsilon^{\alpha \nu \sigma}}{\sqrt{g}} \Bigg[ & \left(  -\R_{\beta \sigma} \nabla^\beta \e^\rho
	+ \R^\rho_{\ \beta} \nabla_\sigma \e^\beta + \nabla_\beta \R^\rho_{\ \sigma} \e^\beta \right) \wedge \delta g_{\nu \rho}     \\
	&+ \delta \R^\rho_{\ \sigma} \wedge \left(    \nabla_\rho \e_\nu + \nabla_\nu \e_\rho     \right)   
	+ \left(    \nabla_\nu \nabla_\beta \e^\rho + R_{\beta\ \nu\gamma}^{\ \rho} \e^\gamma    \right) \wedge \delta \Gamma^\beta_{\ \sigma\rho} \Bigg].
\end{aligned}
\end{equation}

The strategy is to collect terms in the form $ \nabla_\mu X_{CS}^{\mu \alpha} $ plus terms that will cancel the remaining non-boundary terms in the Einstein-Hilbert part (\ref{gauge_einstein}). This can be achieved
by using basic geometric relations, such as the relation between the Riemann tensor and the Einstein tensor in three dimensions, and the identities 
\begin{equation}
\nabla_\beta \delta \R^\beta_{\ \sigma} = \frac{1}{4} \nabla_\sigma \delta R + \delta \Gamma^\lambda_{\ \beta\sigma}  \R^\beta_{\ \lambda}  
		- \delta \Gamma^\lambda_{\ \beta\lambda}  \R^\beta_{\ \sigma},
\end{equation}
\begin{equation}
\epsilon^{\mu \alpha \beta} \e^\nu = g^{\mu \nu} \epsilon^{\rho \alpha \beta} \e_\rho + g^{\alpha \nu} \epsilon^{\mu \rho \beta} \e_\rho 
						+ g^{\beta \nu} \epsilon^{\mu \alpha \rho} \e_\rho.
\end{equation}
After a tedious computation one obtains:
\begin{equation}
\mu \Delta J_{CS}^\alpha = \nabla_\mu  X_{CS}^{\mu \alpha} + C^{\mu \alpha} \left( \e_\mu \wedge \delta \ln g +2  \e^\nu \wedge \delta g_{\mu \nu} \right)
		+ C^{\mu \nu} \delta g_{\mu \nu} \wedge \e^\alpha + 2 \e_\mu \wedge \delta C^{\alpha \mu},
\label{gauge_chern}
\end{equation}
where  the $ X_{CS}^{\mu \alpha}$ is an antisymmetric tensor defined as:
\begin{equation}
X_{CS}^{\mu \alpha} = \frac{\epsilon^{\alpha \mu \sigma}}{\sqrt{g}}  \left(  - \delta\Gamma^\beta_{\ \sigma\rho} \wedge \nabla_\beta \e^\rho
				+ 2 \delta  \R^\nu_{\ \sigma} \wedge \e_\nu +   \R^\rho_{\ \gamma} \delta g_{\sigma \rho}  \wedge \e^\gamma     
				+   \R^\beta_{\ \sigma} \delta g_{\beta \rho}  \wedge \e^\rho      \right).
\label{boundary_chern}
\end{equation}

Combining (\ref{gauge_einstein}) and (\ref{gauge_chern}) and using the field equations and their variations one finds that $\omega$ has no components in the pure gauge directions 
for sufficiently fast decaying metric variations. Finding such a symplectic two-form for TMG was the goal of this paper.

Finally, let us see how conserved charges can be obtained from the above construction. If we restrict the diffeomorphisms to the isometries of the background spacetime, then we have the Killing equation, $ \nabla_\mu \e_\nu + \nabla_\nu \e_\mu = 0$. This leads to 
\begin{equation}
\Delta J^\alpha = \nabla_\mu \left ( X_{EH}^{\mu \alpha} + \frac{1}{\mu} X_{CS}^{\mu \alpha} \right) = 0,
\end{equation}
and to the local conservation $\partial_\mu \left[ \sqrt{g}  \left ( X_{EH}^{\mu \alpha} + \frac{1}{\mu} X_{CS}^{\mu \alpha} \right) \right] = 0$. Strictly speaking, to obtain the conserved charge we should identify
$\delta g_{\mu \nu} \rightarrow h_{\mu \nu}$, where $h_{\mu \nu}$ is a perturbation around a given background with Killing symmetries, 
and keep the $\e^\mu$ terms on the same side of the wedge products before dropping them.
Therefore, the conserved charges can be written (up to a multiplicative constant) as
\begin{equation}
Q^\mu = - \frac{1}{2\pi} \int_{\partial \Sigma} d S_\alpha \sqrt{g}  \left ( X_{EH}^{\mu \alpha} + \frac{1}{\mu} X_{CS}^{\mu \alpha} \right),
\label{charge}
\end{equation}
which gives nonzero results for more slowly decaying metric variations. We adopted the constant $-\frac{1}{2\pi}$ in (\ref{charge}) and the convention that the the one-forms $\e$ to be kept at the right side of the wedge products. With this choice, the conserved charges of TMG read:
\begin{equation}
\begin{aligned}
Q^\mu =  \frac{1}{2\pi} \int_{\partial \Sigma} d S_\alpha \sqrt{g} & \Bigg [ \left(\nabla^\mu h^{\nu \alpha} \e_\nu +  h^{\nu \alpha} \nabla_\nu  \e^\mu  	
				- \frac{1}{2} h  \nabla^\alpha  \e^\mu
		+  \nabla_\nu h^{\mu \nu} \e^\alpha - \nabla^\mu h  \e^\alpha - \left( \alpha \leftrightarrow \mu \right)   \right)  \\
		&+ \frac{1}{\mu}   \frac{\epsilon^{ \mu \alpha \sigma}}{\sqrt{g}}  \left(  - \delta \Gamma^\beta_{\ \sigma\rho} \nabla_\beta \e^\rho
				+ 2 \delta \R^\nu_{\ \sigma} \e_\nu +   \R^\rho_{\ \gamma} h_{\sigma \rho} \e^\gamma     
				+   \R^\beta_{\ \sigma} h_{\beta \rho}  \e^\rho      \right) \Bigg],
\label{charge2}
\end{aligned}
\end{equation}
where $ 2 \delta \Gamma^\beta_{\ \sigma\rho} = g^{\beta \lambda} \left( \nabla_\sigma h_{\rho \lambda} +  \nabla_\rho h_{\sigma \lambda} -  \nabla_\lambda h_{\sigma \rho} \right)$ and 
$\delta \R^\nu_{\ \sigma} = \delta \left(  g^{\nu \lambda} \R_{\lambda \sigma}  \right)$. To compute the latter, one just needs the Palatini identity.
The first line of (\ref{charge2}) is exactly the expression given in \cite{dt}, which can also be recast in the more compact form of \cite{ad}. The second line generalizes the (anti)-de Sitter (AdS) background case of \cite{dt2}. Presumably, the expression given in \cite{cl} for generic backgrounds reduces to the more compact form above. To see that our expression gives the correct charges, we computed the energy of the BTZ black hole \cite{btz} around AdS background. We obtained $E=m-\frac{a}{\mu l^2}$ and $J=a-\frac{m}{\mu}$, which is the same result given in \cite{kanik,olmez}. (Here, $a$ is the rotation parameter in the metric and the vacuum is defined as $a=m=0$.) In the appendix, we consider the conserved charges of three non-Einstein solutions of TMG.

\section{Conclusion}
We have found the symplectic structure of the topologically massive gravity, a closed, conserved, gauge invariant 2-form on the phase space. The nontrivial part of the computation was to show that the symplectic 2-form has vanishing components along the pure diffeomorphism directions. We have also found a compact expression for the conserved Killing charges for generic backgrounds and computed the energy of the BTZ black hole. A covariant canonical quantization can be carried out with the help of the symplectic structure we have presented.

\section{\label{ackno} Acknowledgments}

The work of  B.T. is supported by the TUB\.{I}TAK Grant
No.\ 110T339, and METU Grant No.\ BAP-07-02-2010-00-02.

{\bf  Historical Remark} This work started in 2004  by the suggestion of Yavuz Nutku.  He wrote the first draft and was eager to see its completion but died on  December 7  2010. We dedicate this paper to his memory.

\section{Appendix: Conserved charges for non-Einstein solutions of TMG}
Using the conserved charge expression (\ref{charge2}) let us compute the charges of the previously studied non-Einstein solutions of TMG. The charges of the below metrics have been computed with different techniques before \cite{cl,miskovic,cvetkovic,strominger}.

\subsection{Logarithmic solution of TMG at the chiral point}
At the chiral point $\mu \ell = 1$, where $\ell^2 = -\frac{1}{\Lambda}$, the following metric solves TMG \cite{giribet}:
\begin{equation}
ds^{2}=-N(r)dt^{2}+\frac{dr^{2}}{N(r)}+r^{2}(N_{\theta }(r)dt-d\theta
)^{2}+N_{k}(r)(dt-\ell d\theta)^{2} ,
\end{equation}
where
\begin{equation}
N(r)=\frac{r^{2}}{\ell^{2}}-m+\frac{m^{2}\ell^{2}}{4r^{2}
},\ \   N_{\theta }(r)=\frac{m\ell}{2r^{2}},\ \ 
N_{k}(r)=k\log (\frac{2 r^{2}-m\ell^{2}}{2 r_{0}^{2}}).
\end{equation}
Defining the background as $m=k=0$, our formula (\ref{charge2}) yields the energy (using the Killing vector $\e^\mu = (-1,0,0)$) and the angular momentum (using the Killing vector $\e^\mu = (0,0,1)$) as
\begin{equation}
E=4k,\ \ J=4k\ell.
\end{equation}
These are the same charges as the ones  found in \cite{giribet}, employing the counterterm approach,and in \cite{miskovic}, using the first order formalism, and in \cite{cvetkovic}, employing Nester's definition of conserved charges \cite{nester} 
(Note that here our convention is $8G=1$.)

\subsection {Null warped AdS$_3$}
The following metric solves TMG for  $\mu \ell =-3$  ( See \cite{strominger} and the references therein for the detailed description of the warped AdS metrics )
\begin{equation}
{ds^2\over\ell^2}= -2 r dt d\theta +  \frac{dr^2}{4 r^2} +(r^2+ r+ k) d\theta^2  .
\end{equation}
Taking  $k=0$ case to be the background metric, we compute the charges of this spacetime to  be
\begin{equation}
E =0, \ \ J =   -\frac{ 8 k \ell}{3},
\end{equation}
which are the same as the ones given in \cite{cl,strominger}.

\subsection {Spacelike stretched black holes}

The following metric solves TMG for any value of $\mu$ 
\begin{equation}
ds^2=-N(r) dt^2+\ell^2 R({r})(d\theta+N^\theta(r) dt)^2+{\ell^4d{r}^2\over4
R({r}) N(r)},
\end{equation}
where  the metric functions are given as 
\begin{eqnarray}
R({r})&\equiv&{{r}\over4}\left(3(\nu^2-1){r}+(\nu^2+3)({r}_++{r}_-)-4\nu\sqrt{{r}_+{r}_-(\nu^2+3)}\right),\\
N({r})&\equiv&{\ell^2(\nu^2+3)({r}-{r}_+)({r}-{r}_-)\over4 R({r})}, \ \
N^\theta({r})\equiv{2\nu{r}-\sqrt{{r}_+{r}_-(\nu^2+3)}\over2R({r})},
\end{eqnarray}
where\footnote{Note that with this choice of sign and with the convention $\epsilon^{tr\theta}=1$, the metric solves the TMG equations.} $\nu = -\frac{\mu \ell}{3}$. The solution describes a spacelike stretched black hole for $\nu^2 > 1$ with $r_\pm$ as inner and outer horizons.  This type of solutions to TMG was found by Nutku \cite{nutku} and
 Gurses  \cite{gurses} and studied in \cite{cl,miskovic,cvetkovic,strominger}. The conserved charges of this metric was discussed in the latter works. Using (\ref{charge2}) and defining the background to be $r_\pm =0$ case and using the Killing vectors\footnote{To keep the energy dimensionless we rescale the Killing charge.} $\e^\mu = (-1 / \ell,0,0)$ and $\e^\mu = (0,0,1)$ we get 
\begin{equation}
E=\frac{ \left(3+ \nu^2 \right) }{3 \nu}  \left(   \nu (r_{+} + r_{-} ) - \sqrt{ \left(3+ \nu^2 \right) r_{+} r_{-}}   \right),
\end{equation}
\begin{eqnarray}
J=\frac{\ell}{24\nu} \Bigg[  2 (10 \nu^4 - 15 \nu^2 +9) ( r_{+}^2 &+&  r_{-}^2) + 18 (\nu^2 -1) (\nu^2 -2) r_{+}  r_{-} \\ \nonumber
		&+& \nu (5 \nu^2 -9 ) ( r_{+} +  r_{-})    \sqrt{ \left(3+ \nu^2 \right) r_{+} r_{-}} \Bigg].
\end{eqnarray}

Both $E$ and $J$ turn out to be finite in a highly nontrivial way: Einstein-Hilbert and Chern-Simons parts give divergent results separately, but they yield a finite result when added. Energy computed here is exactly the same as the one given in \cite{cl, cvetkovic, strominger}. However, the angular momentum, $J$, differs from the one, $\mathcal{J}$, given in those papers. $\mathcal{J}$ is a linear combination of $E$ and $J$ given above. The relation is as follows:
\begin{equation}
\mathcal{J} =c_1  J+ c_2 \ell E,
\end{equation}
where
\begin{equation}
c_1 =  \frac{(3+\nu^2)(3+5 \nu^2)}{2 ( \nu^4 + 15 \nu^2 -18)},
\end{equation}
\begin{equation}
c_2 =  -  \frac{     \nu ( 101 \nu^4 - 72 \nu^2  + 27) ( r_{+} +  r_{-})   + 2   \sqrt{ \left(3+ \nu^2 \right) r_{+} r_{-}}  ( 67 \nu^4 + 9 \nu^2  - 72)  }{16 ( \nu^4 + 15 \nu^2 -18)}.
\end{equation}

\end{document}